\documentclass[showkeys,showpacs,prd]{revtex4}
\usepackage{amsmath}
\usepackage{amssymb}
\usepackage{graphicx}
\usepackage{color}

\begin{document}

\title{
Cosmological constant and Euclidean space from nonperturbative quantum torsion
}

\author{Vladimir Dzhunushaliev}
\email{vdzhunus@krsu.edu.kg}
\affiliation{
Department of Theoretical and Nuclear Physics,  Al-Farabi Kazakh National University, Almaty 050040, Kazakhstan, \\
Institute of Experimental and Theoretical Physics,  Al-Farabi Kazakh National University, Almaty 050040, Kazakhstan, \\
Institute for Basic Research,
Eurasian National University,
Astana, 010008, Kazakhstan
}

\begin{abstract}
Heisenberg's nonperturbative quantization technique is applied to the  nonpertrubative quantization of gravity. An infinite set of equations for all Green's functions is obtained. An approximation is considered where: (a) the metric remains as a classical field; (b) the affine connection can be decomposed into classical and quantum parts; (c) the classical part of the affine connection are the Christoffel symbols; (d) the quantum part is the torsion.  Using a scalar and vector fields approximation it is shown that nonperturbative quantum effects gives rise to a cosmological constant and an Euclidean solution.
\end{abstract}


\maketitle

\section{Introduction}

Up to now general relativity has refused to be quantized  -- it remains as an entirely classical theory. Using the same methods for quantization that were successful for electroweak interaction does not bring the desired result.
Let us note that there are similar problems in quantizing quantum chromodynamics e.g. the well known confinement problem. Probably it is not that gravity can not be quantized. There were attempts to quantize gravity by the nonperturbative procedures but the result is non-renormalizable and can in the best case merely be understood as an effective rather than a fundamental theory.

Here we will consider Heisenberg's nonperturbative quantization technique applied to general relativity. In our approach (following  Heisenberg) we write an infinite set of equations for all Green's functions. Probably it is a \emph{causa mortis} to search an exact solution for such an infinite set of equations. More realistically one should use some approximate approach. For example, one could cut off this infinite set of equations using some approximation for n$-th$ Green function. In such an approximation the n$-th$ Green's function is decomposed into the i$-th$ Green's functions with $i<n$. Or one can write some functional generated by the metric and affine connection. Varying with respect to variables describing the Green's functions one can obtain a finite set of equations describing a finite set of Green's functions.

\section{Heisenberg's nonperturbative quantization technique for gravity}
\label{heis}

In 1950's Heisenberg offered a method for the nonperturbative quantization of a nonlinear spinor field \cite{heisenberg}. Following Heisenberg,  nonperturbative operators of a quantum field can be calculated using the corresponding field equation(s) for this theory. \textcolor{blue}{\emph{The main idea is that the corresponding field equation(s) is(are) written in turns of quantum field operators.}}

According to Heisenberg \cite{heisenberg} quantum operators of the metric
$\hat g_{\mu \nu}$ and the affine connection $\hat \Gamma^\rho_{\phantom{\rho} \mu \nu}$ obey the operator Einstein equations in the Palatini formalism
\begin{eqnarray}
	\hat \Gamma^\rho_{\phantom{\rho} \mu \nu} &=&
	\hat G^\rho_{\phantom{\rho} \mu \nu} +
	\hat K^\rho_{\phantom{\rho} \mu \nu} ,
\label{1-10}\\
	\hat R_{\mu \nu} - \frac{1}{2} \hat g_{\mu \nu} \hat R &=& 0 ,
\label{1-20}\\
	\hat G^\rho_{\phantom{\rho} \mu \nu} &=&
	\frac{1}{2} \hat g^{\rho \sigma} \left(
		\frac{\partial \hat g_{\mu \sigma}}{\partial x^\nu} +
		\frac{\partial \hat g_{\nu \sigma}}{\partial x^\mu} -
		\frac{\partial \hat g_{\mu \nu}}{\partial x^\sigma}
	\right)
\label{1-30}
\end{eqnarray}
here we consider Einstein gravity without matter; $\hat R_{\mu \nu}$ is the operator of the Ricci tensor; $\hat G^\rho_{\phantom{\rho} \mu \nu}$ are the operators of the Christoffel symbols; $\hat K^\rho_{\phantom{\rho} \mu \nu}$ is the operator of the contorsion tensor and $\hat R$ is the operator of the scalar curvature defined in usual manner
\begin{eqnarray}
	\hat R_{\mu \nu} &=& \hat R^\rho_{\phantom{\rho} \mu \rho \nu},
\label{1-40}\\
	\hat R^\rho_{\phantom{\rho} \sigma \mu \nu} &=&
	\frac{\partial \hat \Gamma^\rho_{\phantom{\rho} \sigma \nu}}
	{\partial x^\mu} -
	\frac{\partial \hat \Gamma^\rho_{\phantom{\rho} \sigma \mu}}
	{\partial x^\nu} +
	\hat \Gamma^\rho_{\phantom{\rho} \tau \mu}
	\hat \Gamma^\tau_{\phantom{\tau} \sigma \nu} -
	\hat \Gamma^\rho_{\phantom{\rho} \tau \nu}
	\hat \Gamma^\tau_{\phantom{\tau} \sigma \mu} .
\label{1-50}
\end{eqnarray}
The operator of the contortion tensor $\hat K^\rho_{\phantom{\rho} \mu \nu}$ is defined via the operator of the torsion tensor
$\hat Q_{\mu \nu}^{\phantom{\mu \nu}\rho}$
\begin{equation}
	\hat K^\rho_{\phantom{\rho} \mu \nu} =
	\hat Q^\rho_{\phantom{\rho} \mu \nu} +
	\hat Q_{\mu \nu}^{\phantom{\mu \nu} \rho} -
	\hat Q_{\nu \phantom{\rho} \mu}^{\phantom{\mu} \rho}.
\label{1-60}
\end{equation}
The operator of the torsion tensor is the skew-symmetric part of the affine connection
\begin{equation}
	\hat Q_{\mu \nu}^{\phantom{\mu \nu} \rho} = \frac{1}{2} \left(
		\Gamma_{\mu \nu}^{\phantom{\mu \nu} \rho} -
		\Gamma_{\nu \mu}^{\phantom{\mu \nu} \rho}
	\right).
\label{1-70}
\end{equation}
\textcolor{blue}{\emph{The nonperturbative quantization of Einstein gravity means that the quantum  operators $\hat \Gamma, \hat g$ obey the operator Einstein equations \eqref{1-10} \eqref{1-20}.}}

To the end of this section we review the work in Ref. \cite{Dzhunushaliev:2012ca}. Now we have obtained an extremely complicated problem for solving the operator Einstein equations \eqref{1-10} \eqref{1-20}. Heisenberg's approach for solving this problem was to write an infinite set of equations for all Green's functions
\begin{eqnarray}
	\left\langle Q \left| \hat \Gamma(x_1)
	\cdot \text{ Eq. \eqref{1-10} }
	\right| Q \right\rangle &=& 0 ,
\label{1-80}\\
	\left\langle Q \left| \hat g(x_1)
	\cdot \text{ Eq. \eqref{1-10} }
	\right| Q \right\rangle &=& 0 ,
\label{1-90}\\
	\left\langle Q \left| \hat \Gamma(x_1)
	\cdot \text{ Eq. \eqref{1-20} }
	\right| Q \right\rangle &=& 0 ,
\label{1-100}\\
	\left\langle Q \left| \hat g(x_1)
	\cdot \text{ Eq. \eqref{1-20} }
	\right| Q \right\rangle &=& 0 ,
\label{1-110}\\
	\left\langle Q \left| \hat \Gamma(x_1) \hat \Gamma(x_2)
	\cdot \text{ Eq. \eqref{1-10} }
	\right| Q \right\rangle &=& 0 ,
\label{1-120}\\
	\left\langle Q \left| \hat g(x_1) \hat \Gamma(x_2)
	\cdot \text{ Eq. \eqref{1-10} }
	\right| Q \right\rangle &=& 0 ,
\label{1-130}\\
	\left\langle Q \left| \hat g(x_1) \hat g(x_2)
	\cdot \text{ Eq. \eqref{1-10} }
	\right| Q \right\rangle &=& 0 ,
\label{1-140}\\
	\left\langle Q \left| \hat \Gamma(x_1) \hat \Gamma(x_2)
	\cdot \text{ Eq. \eqref{1-20} }
	\right| Q \right\rangle &=& 0 ,
\label{1-150}\\
	\left\langle Q \left| \hat g(x_1) \hat \Gamma(x_2)
	\cdot \text{ Eq. \eqref{1-20} }
	\right| Q \right\rangle &=& 0 ,
\label{1-160}\\
	\left\langle Q \left| \hat g(x_1) \hat g(x_2)
	\cdot \text{ Eq. \eqref{1-20} }
	\right| Q \right\rangle &=& 0 ,
\label{1-170}\\
	\left\langle Q \left| \cdots
	\right| Q \right\rangle &=& 0 ,
\label{1-180}\\
	\left\langle Q \left|
	\text{ the product of $g$ and $\Gamma$ at different points $(x_1,
	\cdots , x_n)$}
	\cdot \text{ Eq. \eqref{1-10} or Eq. \eqref{1-20}}
	\right| Q \right\rangle &=& 0
\label{1-190}
\end{eqnarray}
where $\left. \left|Q \right. \right\rangle$ is a quantum state. For simplicity below we will write $\left\langle \cdots \right\rangle$ instead of
$\left\langle Q \left| \cdots \right| Q \right\rangle$. Schematically the first equation \eqref{1-80} has $\left\langle \Gamma^2 \right\rangle$ and
$\left\langle \Gamma \cdot g \right\rangle$ terms; the second equation \eqref{1-90} has $g \cdot \Gamma$ and $g \cdot g^{\mu \nu} \cdot \partial g$;
the third equation
\eqref{1-100} has $\left\langle \partial\Gamma \cdot \Gamma \right\rangle$ and
$\left\langle \Gamma^3 \right\rangle$ terms and so on up to infinity. Thus all equations are connected and this is the main problem to solve such an infinite set of equations. One can find a similar set of equations in statistical physics and turbulence theory (see, for example, \cite{Wilcox}). Also recall that Heisenberg's matrix mechanics worked with an infinite set of equations in terms of the matrix elements. In perturbative language Eqn.'s \eqref{1-80}-\eqref{1-190} are the Dyson - Schwinger equations but usually (for instance, in quantum electrodynamics) they are written in Feynman diagrams language.

In Eqn.'s \eqref{1-80}-\eqref{1-190} there are products like
$\left. \hat \Gamma^n \right|_{x_i=x}$, $\left. \hat g^n \right|_{x_i=x}$ or $\left. \hat \Gamma^n \right|_{x_i=x} \cdot \left. \hat g^m \right|_{x_i=x}$. By using the perturbative approach (i.e the Feynman diagram technique) these products leads to singularities because the product of field operators for free field, at one point, is poorly defined. We have to point out the difference between such products in perturbative and nonperturbative quantization: for nonperturbative quantization such singularities may be much softer or be absent in general \cite{heisenberg}. The key point is that in the perturbative approach we have moving particles (quanta) only. The communication between two spacetime points is carried by such quanta. Consequently the correlation function between these points (Green's functions) is non-zero only if the interaction between points via quanta is possible. This means that the corresponding Green's function is non-zero inside of the light cone and zero outside the light cone. Such functions have to be singular on the light cone. But for the nonperturbative case this is not true: for  self-interacting, nonlinear fields there may exist static configurations. For instance, it can be a nucleon, a glueball, a flux tube stretched between quark and antiquark, filled with a longitudinal chromoelectric field, etc.  In this case the correlation function (Green's function) is non-zero outside of the light cone. This is the reason why the Green's function for the nonperturbative case is not so singular as compared with the perturbative case.

Mathematically it can be explained as follows: for free and interacting operators of quantum fields we have two \textcolor{blue}{\emph{different}}  algebras. For free fields the algebra is well known and it is described by canonical commutations relationships where commutators and anticommutators are distributions. But for interacting fields the algebra is unknown and in the arguments above we put forward the idea that the defining relationships will be ordinary functions (not distributions).

It is highly probable that the set of equations -- either \eqref{1-10}-\eqref{1-30}  or \eqref{1-80}-\eqref{1-190} --
can not be solved analytically. One possible way to approximately solve these equations is as follows: We decompose the n$-th$ Green's function
\begin{equation}
	G_n = G(x_1, x_2 \cdots , x_n) =
	\left\langle
		\Gamma(x_1) \cdots g(x_m) \cdots
	\right\rangle
\label{1-200}
\end{equation}
into a linear combination of products of Green's functions of lower orders
\begin{equation}
\begin{split}
	G_n(x_1, x_2 \cdots , x_n) \approx & G_{n-2}(x_3, x_4 \cdots , x_n)
	\left[ G_2(x_1, x_2) -C_2 \right] +
\\
	&
	\left( \text{permutations of $x_1,x_2$ with
	$x_3, \cdots , x_n$} \right) +
\\
	&	
	G_{n-3} \left( G_3 -C_3 \right) + \cdots
\end{split}
\label{1-210}
\end{equation}
where $C_{2,3 \cdots}$ are constants. In such a way one can cut off the infinite set of equations \eqref{1-80}-\eqref{1-190}. Such a technique is used to solve similar sets of equations in statistical physics and turbulence theory.

Another way to approximately solve the infinite set of equation \eqref{1-80}-\eqref{1-190} is to choose some functional (for instance, the action or something like the gluon condensates in quantum chromodynamics \footnote{in quantum chromodynamics there exist various gluon condensates:
$\mathrm{tr}\left(F_{\mu \nu} F^{\mu \nu} \right)$,
$\mathrm{tr} \left( F_\mu^\nu F_\nu^\rho F_\rho^\mu \right)$ and so on. For a review on gluon condensate one can see Ref. \cite{Zakharov:1999jj}. Integrating these condensates one can obtain various functionals.}) and write down its average expression using the corresponding Green's functions. After this it is necessary to use some well-reasoned physical assumptions to express the highest order Green's function through Green's functions of lower orders e.g. see the decomposition \eqref{1-210}. Finally we will have a functional which can be used to obtain the Euler - Lagrange field equations for the Green's functions.

Let us note that equations \eqref{1-80}-\eqref{1-190} are similar to: (a) a Bogolyubov chain of equations (hierarchy) for the one-particle, two-particle, etc., distribution functions of a classical statistical system; (b) Dyson - Schwinger equations for n-point Green's functions in quantum field theory; (c) equations connecting correlation functions for velocities, pressure and density in a turbulent fluid. In all cases the solution methods are similar to those discussed above: the decomposition of the n-point corresponding functions in terms of the product of lower order functions. For us the more interesting case is (b), the Dyson - Schwinger equations. In a perturbative quantum field theory these equations are usually written in the language of Feynman diagram techniques. But for us this approach is not applicable because gravity is a strongly self-interacting system. Heisenberg applied nonperturbative techniques for the quantization of a nonlinear, spinor field \cite{heisenberg}. In Ref. \cite{Dzhunushaliev:2010qs} Heisenberg's technique was applied to the quantization of nonpertubative quantum chromodynamics, and it was shown that such an approach leads to the description of a gluon condensate in: (a) a glueball; (b) a flux tube filled with a longitudinal color electric field and stretched between quark - antiquark located at $\pm$ infinities. Following this way in Ref. \cite{Dzhunushaliev:2002ve} was shown that London'd equations can be obtained from a non-Abelian gauge theory.

The approximations from \cite{heisenberg, Dzhunushaliev:2010qs} are not applicable for the quantization of general relativity because the nonlinearities in gravity are extremely strong. In usual field theory the nonlinearities are in the potential terms but in gravity they are in the kinetic term as well. Therefore we can not apply the nonperturbative quantization methods from usual quantum field theory to the quantization of gravity.

\section{Scalar field approximation for Einstein equations corrected by quantum torsion}
\label{sec3}

In this section we will show that in quantum gravity fluctuating torsion gives rise to a cosmological constant both by perturbative and nonperturbative calculations in quantum gravity. In the approach presented here we consider the case with a classical metric, $g_{\mu \nu}$, and quantum affine connection, $\hat \Gamma = G + \hat K$, where the Christoffel symbols (denoted by  $G$) are in a classical mode and the contorsion tensor $\hat K$ is in a quantum mode.

Firstly we recall some relevant results from perturbative calculations \cite{ivanenko}. Then we show that these same result can be obtained via nonperturbative quantum gravity.

\subsection{Perturbative calculations with fluctuating torsion}

The torsion contribution in perturbative quantum gravity is quite well studied. In Ref. \cite{Elizalde:1992nc} the path integral for higher-derivative quantum gravity with torsion is considered. The calculations are made in the limit of conformally self-dual metrics. The authors found that torsion actually shows up in a very complicated way, and that it cannot be integrated over. In Ref. \cite{Elizalde:1993uc} the general expression describing the conformal dynamics of quantum gravity with torsion in a curved fiducial background is obtained. The one-loop effective potential for the conformal factor (up to terms linear on the curvature and up to second order in torsion) is calculated. In Ref. \cite{Bytsenko:1993qn} the authors apply the renormalization group approach to the calculation and analysis of the effective potential corresponding to interacting quantum field theory in curved spacetime with torsion. The purpose of Ref. \cite{Antoniadis:1992ep} is the investigation of the trace anomaly induced dynamics of the conformal factor in four-dimensional quantum gravity.
In Ref. \cite{Buchbinder:1989sm} Grand Unification Theories in curved spacetime with torsion are investigated. It is shown that in a strong gravitational field these effective coupling constants tend to conformal values or increase in an exponential way.

In Ref. \cite{ivanenko} the torsion was considered as a collection of classical fields that should be quantized. This collective classical field could be written as
\begin{equation}
	Q_\mu = Q_\mu^0 + q_\mu
\label{3a-10}
\end{equation}
where $Q_\mu^0$ is the classical torsion and $q_\mu$ are quantum fluctuations. Minimizing the effective potential of the torsion field and taking into account the polarization of a matter field on the vacuum it was shown that
\begin{equation}
	Q_\mu^0 = 0.
\label{3a-20}
\end{equation}
The calculations further show that
\begin{equation}
	\left\langle  Q_\mu Q^\mu \right\rangle = 2 \Delta_0^2 +
	\text{radiative corrections}
\label{3a-30}
\end{equation}
where $\Delta_0^2$ is some constant. The result is that at low energies, and near an energy minimum, that gravitational theory with torsion is equivalent to Einstein gravity with a cosmological constant. The cosmological constant is the consequence of a spin-spin gravitational interaction.

In the next subsection we would like to show that a similar result can be obtained by a nonperturbative quantization of gravity.

\subsection{Nonperturbative calculations with fluctuating torsion}

Now we want to emphasize once again that, since quantum gravity is a non-renormalizable theory, the quantization of gravity must be performed without using the Feynman diagram technique. One way of performing this procedure is Heisenberg's nonperturbative quantization technique presented in section \ref{heis}. The main physical difference between Heisenberg and Feynman approaches is that Feynman diagram technique is based on the fact that the interaction between fields occurs at the points (vertexes). The quanta move between vertexes as free particles. In contrast to Feynman approach the Heisenberg approach is based on the fact that \emph{the interactions between quantum fields occur at all points of spacetime}. Mathematically it means that instead of drawing and calculations of all Feynman diagrams we have to solve infinite equations set \eqref{1-80}-\eqref{1-190}. Practically we cannot solve infinite equations set \eqref{1-80}-\eqref{1-190} and we have to cut off such infinite equations set to finite equations set using some physical assumptions about Green functions. For example, it can be so called either scalar \eqref{3b-60} or vector \eqref{4-40} approximation for 2-point Green function.

For the derivation of the cosmological constant as a quantum gravity nonperturbative effect we will assume that the expectation value of the contortion tensor is zero
\begin{equation}
	\left\langle
		\hat \Gamma^\rho_{\phantom{\rho} \mu \nu}
	\right\rangle =
	G^\rho_{\phantom{\rho} \mu \nu}
\label{3b-10}
\end{equation}
where $G^\rho_{\phantom{\rho} \mu \nu}$ are the classical Christoffel symbols. This means that
\begin{equation}
	\left\langle
		\hat K^\rho_{\phantom{\rho} \mu \nu}
	\right\rangle = 0
\label{3b-20}
\end{equation}
but the standard deviation of the contortion tensor is not zero
\begin{equation}
	\left\langle
		\left( \hat \Gamma^\rho_{\phantom{\rho} \mu \nu} \right)^2
	\right\rangle \neq
	\left( G^\rho_{\phantom{\rho} \mu \nu} \right)^2
\label{3b-30}
\end{equation}
i.e.
\begin{equation}
	\left\langle
		\left( \hat K^\rho_{\phantom{\rho} \mu \nu} \right)^2
	\right\rangle \neq 0.
\label{3b-40}
\end{equation}
For simplicity we also assume that the contortion tensor is absolutely antisymmetric i.e.
$\hat K_{\rho \mu \nu} = \hat K_{[\rho \mu \nu]}$. In this case
\begin{equation}
	\hat K_{\rho \mu \nu} = \hat Q_{\rho \mu \nu} = \hat Q_{[\rho \mu \nu]}
\label{3b-50}
\end{equation}
where $\hat Q_{\rho \mu \nu}$ is the torsion tensor operator. Our strategy for the nonperturbative calculations is the following:
\textit{
\textcolor{blue}{
using some assumptions about the standard deviation of the torsion tensor such as
$
\left\langle
		\left( \hat Q^\rho_{\phantom{\rho} \mu \nu} \right)^2
\right\rangle
$
we will average the Einstein operator equation \eqref{1-20} with the operator of the affine connection from \eqref{1-10} and the operator of the Ricci tensor from \eqref{1-40} \eqref{1-50}.
}
}

Our basic assumption is the so called \emph{scalar field approximation}
\begin{equation}
	\left\langle
		\hat Q_{\rho_1 \mu_1 \nu_1}(x) \hat Q_{\rho_2 \mu_2 \nu_2}(x)
	\right\rangle \approx \varsigma
	\varepsilon_{\rho_1 \mu_1 \nu_1 \alpha}(x)
	\varepsilon_{\rho_2 \mu_2 \nu_2 \beta}(x)
	g^{\alpha \beta}(x)
	\left| \phi(x) \right|^2
\label{3b-60}
\end{equation}
where $\varepsilon_{\rho \mu \nu \sigma}$ is the absolutely antisymmetric  Levi - Civita tensor; $\varsigma = \pm 1$ \footnote{Unfortunately our approach can not give us the sign of the term appearing from the quantum torsion fluctuations} and $\phi(x)$ is a scalar field.

According to \cite{poplawski} the relation between the Ricci tensor $R_{\mu \nu}$ obtained from the affine connection $\Gamma^\rho_{\phantom{\rho} \mu \nu}$ and the Ricci tensor
$\tilde R_{\mu \nu}$ obtained from the Christoffel symbols $G^\rho_{\phantom{\rho} \mu \nu}$ is the following
\begin{equation}
  R_{\mu \nu} = \tilde R_{\mu \nu} +
  Q^{\alpha}_{\mu \nu ; \alpha} -
  Q^{\alpha}_{\mu \alpha ; \nu} +
  Q^\rho_{\phantom{\rho} \mu \nu}
  Q^\sigma_{\phantom{\sigma} \rho \sigma} -
	Q^\rho_{\phantom{\rho} \mu \sigma}
	Q^\sigma_{\phantom{\sigma} \rho \nu}.
\label{3b-63}
\end{equation}
Corresponding expressions for the operators are
\begin{equation}
  \hat R_{\mu \nu} = \hat{\tilde R}_{\mu \nu} +
  \hat Q^{\alpha}_{\mu \nu ; \alpha} -
  \hat Q^{\alpha}_{\mu \alpha ; \nu} +
  \hat Q^\rho_{\phantom{\rho} \mu \nu}
  \hat Q^\sigma_{\phantom{\sigma} \rho \sigma} -
	\hat Q^\rho_{\phantom{\rho} \mu \sigma}
	\hat Q^\sigma_{\phantom{\sigma} \rho \nu}.
\label{3b-66}
\end{equation}
Averaging of the Ricci tensor operator \eqref{3b-66} gives us
\begin{equation}
	\left\langle
		\hat R_{\mu \nu}(x)
	\right\rangle = \tilde{R}_{\mu \nu}(x) + \left\langle
		\hat Q^\rho_{\phantom{\rho} \mu \nu}(x)
		\hat Q^\sigma_{\phantom{\sigma} \rho \sigma}(x)
	\right\rangle - \left\langle
		\hat Q^\rho_{\phantom{\rho} \mu \sigma}(x)
		\hat Q^\sigma_{\phantom{\sigma} \rho \nu}(x)
	\right\rangle
\label{3b-70}
\end{equation}
where we took into account that
$\left\langle	\hat K^\rho_{\phantom{\rho} \mu \nu}
\right\rangle = \left\langle	\hat Q^\rho_{\phantom{\rho} \mu \nu}
\right\rangle = 0$ and
$
\left\langle
  \hat{\tilde R}_{\mu \nu}
\right\rangle = \tilde{R}_{\mu \nu}
$ in the consequence of the fact that we are considering the case with a classical metric (as mentioned above at the beginninig of section \ref{sec3}). Due to the antisymmetry of the torsion \eqref{3b-50} we have $\hat Q^\sigma_{\phantom{\sigma} \rho \sigma} = 0$ and the expectation value of the Ricci operator is
\begin{equation}
	\left\langle
		\hat R_{\mu \nu}(x)
	\right\rangle = \tilde{R}_{\mu \nu}(x) -
	6 \varsigma g_{\mu \nu} \left| \phi \right|^2.
\label{3b-80}
\end{equation}
Now we can calculate the expectation value of the left side of the Einstein equations
\begin{equation}
	\left\langle
		\hat R_{\mu \nu}(x) - \frac{1}{2} g_{\mu \nu} \hat R
	\right\rangle = \tilde{R}_{\mu \nu}(x) - \frac{1}{2} g_{\mu \nu} \tilde R -
	6 \varsigma g_{\mu \nu} \left| \phi \right|^2.
\label{3b-90}
\end{equation}
Thus the vacuum Einstein equations with nonperturbative quantum gravitational admixtures are
\begin{equation}
	\tilde{R}_{\mu \nu} - \frac{1}{2} g_{\mu \nu} \tilde R -
	6 \varsigma g_{\mu \nu} \left| \phi \right|^2 = 0
\label{3b-100}
\end{equation}
Now we would like to obtain an equation for the scalar field $\phi$ approximately describing nonperturbative quantum gravitational effects. In order that vacuum Einstein equations
\begin{equation}
	\left\langle
		\hat{R}_{\mu \nu} - \frac{1}{2} g_{\mu \nu} \hat R
	\right\rangle = 0
\label{3b-105}
\end{equation}
are not overdetermined we demand that
\begin{equation}
	\left\langle
	\left(
		\hat{R}^\mu_\nu - \frac{1}{2} \delta^\mu_\nu \hat R
	\right)_{; \mu}
	\right\rangle = 0
\label{3b-107}
\end{equation}
Taking into account that
\begin{equation}
	\left( \tilde{R}^\mu_\nu - \frac{1}{2} \delta^\mu_\nu
	\tilde R \right)_{;\mu}	= 0
\label{3b-110}
\end{equation}
we obtain the desired equation for $\phi$
\begin{equation}
	\left| \phi \right|_{;\mu} = \left| \phi \right|_{, \mu} = 0 .
\label{3b-130}
\end{equation}
Eq. \eqref{3b-110} is the Bianchi identity for the quantities
$\tilde{R}_{\mu \nu}$ and $\tilde R$ constructed from the Christoffel symbols.
The $\phi$-equation \eqref{3b-130} gives us following solution
\begin{equation}
	\phi = \mathrm{const}
\label{3b-140}
\end{equation}
and we can identify the scalar field describing the nonperturbative quantum effects with a cosmological constant in the following way
\begin{equation}
	\Lambda = - 6 \varsigma \left| \phi \right|^2.
\label{3b-150}
\end{equation}
The results of this subsection are following:
\begin{itemize}
	\item nonperturbative quantum gravitational effects lead to a cosmological constant;
	\item these nonperturbative effects appear from quantum torsion.
\end{itemize}

\section{Euclidean metric from nonperturbative quantum gravity effects}

In the previous section we have shown that in the first approximation nonperturbative quantum gravity effects lead to the appearance of a cosmological constant. Now we would like to consider another approach: \emph{a vector field approximation} for $\left\langle \hat Q^2 \right\rangle$. We will see that in this case nonperturbative quantum gravity effects give rise to an Euclidean metric. Following Hawking \cite{Hawking} a small part of this space can be considered as a Euclidean origin of our Universe that should be joined (with the change of metric signature) to a Lorentzian spacetime.

\subsection{Vector field approximation for Einstein equations corrected by fluctuating torsion}

The main difference between this section and the previous section is the assumption about the expectation value of the square of torsion operator. Unlike the scalar field approximation \eqref{3b-60} we will use in this section a vector field approximation
\begin{eqnarray}
  \hat K_{\rho \mu \nu} &=& \hat Q_{\rho \mu \nu} = \hat Q_{[\rho \mu \nu]}.
\label{4-10}\\
	\left\langle
		\hat Q^\rho_{\phantom{\rho} \mu \nu}
	\right\rangle &=& 0
\label{4-20}\\
	\left\langle
		\left( \hat Q^\rho_{\phantom{\rho} \mu \nu} \right)^2
	\right\rangle &\neq& 0.
\label{4-30}\\
	\left\langle
		\hat Q_{\rho_1 \mu_1 \nu_1}(x_1) \hat Q_{\rho_2 \mu_2 \nu_2}(x_2)
	\right\rangle &\approx& \varsigma \;
  \varepsilon_{\rho_1 \mu_1 \nu_1 \alpha}(x_1)
	\varepsilon_{\rho_2 \mu_2 \nu_2 \beta}(x_2)
	A^\alpha (x_1) A^\beta (x_2)
\label{4-40}
\end{eqnarray}
where we again introduce $\varsigma = \pm 1$ since we can not fix the sign of the RHS of \eqref{4-40}. The expectation values of the Ricci and scalar curvature operators with the vector field approximation for nonperturbative gravity quantization are
\begin{eqnarray}
	\left\langle
		\hat R_{\mu \nu}
	\right\rangle &=& \tilde{R}_{\mu \nu} -
	\left\langle
		\hat Q^\rho_{\phantom{\rho} \mu \sigma}
		\hat Q^\sigma_{\phantom{\sigma} \rho \nu}
	\right\rangle = \tilde{R}_{\mu \nu} -
	\varsigma \; \varepsilon^\rho_{\phantom{\rho} \mu \sigma \alpha}
	\varepsilon^\sigma_{\phantom{\sigma}\rho \nu \beta}
	A^\alpha A^\beta = \tilde{R}_{\mu \nu} + 2 \varsigma \left(
		g_{\mu \nu} A_\alpha A^\alpha - A_\mu A_\nu
	\right) ,
\label{4-50}\\
	\left\langle	\hat R	\right\rangle &=&
	\tilde{R} + 6 \varsigma A^\mu A_\mu .
\label{4-60}
\end{eqnarray}
The vacuum Einstein equations are
\begin{equation}
	\tilde{R}_{\mu \nu} -
		\frac{1}{2} g_{\mu \nu} \tilde{R} - \varsigma \left(
		g_{\mu \nu} A_\alpha A^\alpha + 2 A_\mu A_\nu
	\right) = 0
\label{4-70}
\end{equation}
The same requirement as \eqref{3b-105}-\eqref{3b-110} gives us the following equation for the vector field $A_\mu $
\begin{equation}
	\left(
		\delta^\mu_\nu A^\alpha A_\alpha + 2 A^\mu A_\nu
	\right)_{; \mu} = 0
\label{4-80}
\end{equation}
or
\begin{equation}
	\left( A^\alpha A_\alpha \right)_{, \nu} + 2 A^\mu_{; \mu} A_\nu +
	2 A^\mu A_{\nu ; \mu} = 0
\label{4-90}
\end{equation}

\subsection{Euclidean solution}

In this subsection we solve Einstein equations \eqref{4-70} with terms describing nonperturbative quantum corrections appearing from the quantum fluctuating torsion. These corrections are approximately described with a vector field $A_\mu$ obeying Eq. \eqref{4-90}. Our main goal here is to show that near the origin (where a singularity is located) the metric (which is the solution of equations \eqref{4-70} \eqref{4-80}) is regular. In other words the nonperturbative quantum gravity effects do away with the singularities.

Let us consider the spherically symmetric metric
\begin{equation}
	ds^2 = b^2(r) \Delta(r) dt^2 -
	\frac{dr^2}{\Delta(r)} -
	r^2 \left(
		d \theta^2 + \sin^2 \theta d \varphi^2
	\right)
\label{4b-10}
\end{equation}
and the vector $A_\mu$ is
\begin{equation}
	A_\mu = \left( \phi(r), 0, 0, 0 \right).
\label{4b-15}
\end{equation}
The corresponding Einstein and $A_\mu$ equations \eqref{4-80} are
\begin{eqnarray}
	- \frac{\Delta'}{\Delta} -
	\frac{\Delta - 1}{r \Delta} -
	3 \varsigma \frac{r \phi^2}{b^2 \Delta^2}  &=& 0,
\label{4b-20}\\
	2 \frac{b'}{b} + \frac{\Delta'}{\Delta} + \frac{\Delta - 1}{r \Delta} +
	\varsigma \frac{r \phi^2}{b^2 \Delta^2} &=& 0,
\label{4b-30}\\
	\frac{1}{2} \frac{\Delta''}{\Delta} + \frac{b''}{b} +
	\frac{3}{2}\frac{b' \Delta'}{b \Delta} +
	\frac{\Delta'}{r \Delta} + \frac{b'}{r b} +
	\varsigma \frac{\phi^2}{b^2 \Delta^2} &=& 0,
\label{4b-40}\\
	\frac{\phi'}{\phi} - \frac{2 b'}{b} - \frac{\Delta'}{\Delta} &=& 0.
\label{4b-50}
\end{eqnarray}
The solution of this equations set is given in the Appendix \ref{app1}.

We will consider the solution \eqref{app-90} with $C_1 = 0$
\begin{eqnarray}
	b^2 &=& - 	\frac{1}{C + \varsigma \frac{r^2}{l_0^2}},
\label{4b-60}\\
	\Delta &=& \frac{1}{C}
  \left( C + \varsigma \frac{r^2}{r_0^2} \right).
\label{4b-70}
\end{eqnarray}
The metric \eqref{4b-10} is
\begin{equation}
	ds^2 = - \frac{1}{C} dt^2 -
	\frac{dr^2}{1 + \frac{\varsigma}{C} \frac{r^2}{r_0^2}} -
	r^2 \left(
		d \theta^2 + \sin^2 \theta d \varphi^2
	\right)
\label{4b-80}.
\end{equation}
Considering the case $\varsigma , C = +1$ and introducing new coordinate $\chi = \mathrm{arssinh} \frac{r}{l_0}$ we obtain the metric
\begin{equation}
	ds^2 = - dt^2 - l_0^2 \left[
	d\chi^2 + \sinh^2 \chi \left(
		d \theta^2 + \sin^2 \theta d \varphi^2
	\right)
	\right]
\label{4b-90}.
\end{equation}
and we obtain a very unexpected result: $g_{tt} = - 1$ is negative and $t$ becomes an imaginary time. In our opinion the physical sense of this result is the following: \textcolor{blue}{\emph{the nonperturbative quantum gravitational corrections gives rise to solutions where the signature of the spacetime depends on an arbitrary parameter}} (constant C in our case). We note that the scalar curvature $\tilde R$ is nonsingular.

Following Hawking \cite{Hawking} one can offer the following interpretation of this solution: quantum effects are essential near the origin, $t=0$, only. Far away from the origin these effects can be neglected. This allows us to assume that far away from the origin we have to join the metric \eqref{4b-90} with a cosmological metric having a Big Bang singularity. This means that the nonperturbative quantum gravitational corrections may replace the cosmological singularity with a transition from an Euclidean to a Lorentzian metric.

\section{Conclusions}

We have shown that Heisenberg's nonperturbative quantization technique can be applied to quantum gravity. We have written an infinite set of equations describing all Green functions for the metric and the affine connection. Such a system of equations can not be solved exactly and we have offered some approaches to solve approximately this system.

In this paper we have considered an approximation (for the Palatini formalism) when the metric remains as a classical field and the affine connection as a quantum field. In our approximation the affine connection can be split into two parts: the first one is the classical Christoffel symbols and the second one is the quantum torsion. Using a scalar and vector field approximations we have shown that: (a) in the first case nonperturbative quantum corrections looks like a cosmological constant; (b) in the second case nonperturbative quantum corrections gives rise to a Euclidean metric  near the origin. Formally we have the interesting possibility to interpret integration constant $C$ (see metric \eqref{4b-80}) as the metric signature: for $C=+1$ we have regular Euclidean space and for $C=-1$ we have Lorentzian spacetime.
\par
Another interesting result is that the vacuum Einstein equations gives us regular solutions. The reason for this fact is that nonperturbative quantum gravitational effects are taken into account.

\section*{Acknowledgments}

I am grateful to the Volkswagen Foundation. This work was partially supported  by grants \#1626/GF3 and \#139 in fundamental research in natural sciences by the Science Committee of the Ministry of Education and Science of Kazakhstan. I am grateful to D. Singleton for fruitful remarks and comments.

\appendix

\section{Solution of corrected Einstein equations in the vector field approximation}
\label{app1}

Equation \eqref{4b-50} has the following solution
\begin{equation}
	\phi = \frac{b^2\Delta}{l_0}
\label{app-10}
\end{equation}
where $l_0$ is an integration constant. That leads to the following set of equations
\begin{eqnarray}
	\frac{\Delta'}{\Delta} +
	\frac{\Delta - 1}{r \Delta} + 3 \varsigma \frac{r b^2}{l_0^2}  &=& 0,
\label{app-20}\\
	\frac{b'}{b} &=& \varsigma \frac{r b^2}{l_0^2},
\label{app-30}\\
	\frac{1}{2} \frac{\Delta''}{\Delta} + \frac{b''}{b} +
	\frac{3}{2}\frac{b' \Delta'}{b \Delta} +
	\frac{\Delta'}{r \Delta} + \frac{b'}{r b} +
	\varsigma \frac{b^2}{l_0^2} &=& 0.
\label{app-40}
\end{eqnarray}
Equation \eqref{app-30} has the solution
\begin{equation}
	b^2 = - \frac{1}{C + \varsigma \frac{r^2}{l_0^2}}
\label{app-50}
\end{equation}
where $C$ is an integration constant.

Let us consider two different cases.

\textbf{1. $C=0$.} In this case taking into account equation \eqref{app-20} we have the following solution
\begin{eqnarray}
	b &=& \sqrt{- \varsigma} \; \frac{l_0}{r},
\label{app-60}\\
	\Delta &=& \frac{1}{2} \left(
		\frac{r^2}{r_0^2} - 1
	\right)
\label{app-70}
\end{eqnarray}
where $r_0$ is a constant. The metric has the form
\begin{equation}
	ds^2 = - \frac{\varsigma}{2} \left(
		1 - \frac{l_0^2}{r^2}
	\right) dt^2 - \frac{2 dr^2}{\frac{r^2}{l_0^2} - 1} -
	r^2 \left(
		d \theta^2 + \sin^2 \theta d \varphi^2
	\right)
\label{app-80}
\end{equation}
where we have rescaled the time $t$ in such a way that $r_0 = l_0$.

\textbf{2. $C \neq 0$.} In this case equation \eqref{app-20} gives us the following solution
\begin{equation}
	\Delta = 1 + \frac{\varsigma}{C} \frac{r^2}{l_0^2} + C_1
	\frac{l_0}{r} \left( C + \varsigma \frac{r^2}{l_0^2} \right)^{3/2} =
   \left( C + \varsigma \frac{r^2}{l_0^2} \right)^{1/2}
   \left[
    \frac{1}{C} + C_1
    \left( \varsigma + C \frac{l_0^2}{r^2} \right)^{1/2}
   \right]
\label{app-90}
\end{equation}
where $C_1$ is an integration constant. The metric has the form
\begin{equation}
	ds^2 = - \left[
		\frac{1}{C} +
		C_1
		\left( \varsigma + C \frac{l_0^2}{r^2} \right)^{1/2}
	\right] dt^2 -
	\frac{dr^2}{\left(
		C + \varsigma \frac{r^2}{l_0^2}
	\right)
	\left[
		\frac{1}{C} + C_1 \left(
			\varsigma + C \frac{l_0^2}{r^2}
		\right)^{1/2}
	\right]} -
	r^2 \left(
		d \theta^2 + \sin^2 \theta d \varphi^2
	\right)
\label{app-95}
\end{equation}

\end{document}